\documentclass[sn-mathphys-num]{sn-jnl}

\usepackage{graphicx}%
\usepackage{multirow}%
\usepackage{amsmath,amssymb,amsfonts}%
\usepackage{amsthm}%
\usepackage{mathrsfs}%
\usepackage[title]{appendix}%
\usepackage{xcolor}%
\usepackage{textcomp}%
\usepackage{manyfoot}%
\usepackage{booktabs}%
\usepackage{algorithm}%
\usepackage{algorithmicx}%
\usepackage{algpseudocode}%
\usepackage{listings}%
\usepackage{soul}

\theoremstyle{thmstyleone}%
\theoremstyle{thmstyletwo}%

\theoremstyle{thmstylethree}%

\begin{document}

\title{Agreement of Image Quality Metrics with Radiological Evaluation in the Presence of Motion Artifacts}


\author[1,2,3]{\fnm{Elisa} \sur{Marchetto}} \equalcont{These authors contributed equally to this work.}
\author[4,5]{\fnm{Hannah} \sur{Eichhorn}} \equalcont{These authors contributed equally to this work.}
\author[3]{\fnm{Daniel} \sur{Gallichan}}
\author[4,5,6]{\fnm{Julia A.} \sur{Schnabel}}
\author*[7,8]{\fnm{Melanie} \sur{Ganz}}\email{melanie.ganz@nru.dk}

\affil[1]{\orgdiv{Bernard and Irene Schwartz Center for Biomedical Imaging, Dept. of Radiology}, \orgname{NYU School of Medicine}, \orgaddress{\state{NY}, \country{USA}}}
\affil[2]{\orgdiv{Center for Advanced Imaging Innovation and Research (CAI\textsuperscript{2}R), Dept. of Radiology}, \orgname{NYU School of Medicine}, \orgaddress{\state{NY}, \country{USA}}}
\affil[3]{\orgdiv{CUBRIC, School of Engineering}, \orgname{Cardiff University}, \orgaddress{\state{Cardiff}, \country{UK}}}
\affil[4]{\orgdiv{Institute of Machine Learning in Biomedical Imaging}, \orgname{Helmholtz Munich}, \orgaddress{\state{Neuherberg}, \country{Germany}}}
\affil[5]{\orgdiv{School of Computation, Information and Technology}, \orgname{Technical University of Munich}, \orgaddress{\state{Munich}, \country{Germany}}}
\affil[6]{\orgdiv{School of Biomedical Engineering and Imaging Sciences}, \orgname{King's College London}, \orgaddress{\state{London}, \country{UK}}}
\affil[7]{\orgdiv{Department of Computer Science}, \orgname{University of Copenhagen}, \orgaddress{\state{Copenhagen}, \country{Denmark}}}
\affil[8]{\orgdiv{Neurobiology Research Unit}, \orgname{Copenhagen University Hospital}, \orgaddress{\state{Copenhagen}, \country{Denmark}}}


\abstract{\textbf{Object:} Reliable image quality assessment is crucial for evaluating new motion correction methods for magnetic resonance imaging. We compare the performance of common reference-based and reference-free image quality metrics on unique datasets with real motion artifacts, and analyze the metrics' robustness to typical pre-processing techniques.

\textbf{Materials and Methods:} We compared five reference-based and five reference-free metrics on brain data acquired with and without intentional motion (2D and 3D sequences). The metrics were recalculated seven times with varying pre-processing steps. Spearman correlation coefficients were computed to assess the relationship between image quality metrics and radiological evaluation.

\textbf{Results:} All reference-based metrics showed strong correlation with observer assessments. Among reference-free metrics, Average Edge Strength offers the most promising results, as it consistently displayed stronger correlations across all sequences compared to the other reference-free metrics. The strongest correlation was achieved with percentile normalization and restricting the metric values to the skull-stripped brain region. In contrast, correlations were weaker when not applying any brain mask and using min-max or no normalization.

\textbf{Discussion:} Reference-based metrics reliably correlate with radiological evaluation across different sequences and datasets. Pre-processing significantly influences correlation values. Future research should focus on refining pre-processing techniques and exploring approaches for automated image quality evaluation.
}




\maketitle

\section{Introduction}\label{sec:intro}

Quantitative evaluation of image quality is crucial across various sub-fields of magnetic resonance imaging (MRI). Particularly, the development and thorough validation of new image reconstruction and artifact correction techniques requires reliable quantitative image quality assessment. A large number of image quality metrics (IQMs) are  employed in the literature, with some being \textit{reference-based} metrics that require a ground truth or reference image, and others being \textit{reference-free} \cite{tisdall2022metrics}.\footnote{To avoid confusion we note that sometimes reference-based metrics are also referred to as paired metrics and reference-free metrics as unpaired metrics.} However, none of these metrics are sensitive to all types of image artifacts and the lack of standardized image quality evaluation might lead to "metric-picking". Thus, the use of IQMs for benchmarking different image reconstruction or motion correction methods is challenging and might misguide future research \cite{Heckel_2024,Spieker_2024,breger2024study}.

Most IQMs were originally designed for natural images and their performance in the medical domain may not yet have been thoroughly validated \cite{Heckel_2024,breger2024study}. Medical image quality can be defined as how well the desired clinical information, i.e. the clinical diagnosis, can be extracted from the image in the relevant downstream task \cite{Barrett_1993}. In practice, however, reference values for task-based quality measures are challenging to define and time consuming to obtain. Hence, radiological evaluation of overall image quality is commonly used as a \textit{gold standard} when investigating the performance of IQMs \cite{Mason_2020,Kastryulin_2023,Eichhorn_2022,Marchetto_2024}. Such an evaluation is typically based on the radiologist's assessment of signal-to-noise ratio, sharpness, blurring and presence of artefacts in the images.

In the context of MR image reconstruction and motion correction, structural similarity index (SSIM) and peak signal-to-noise ratio (PSNR) are among the most commonly used IQMs. Yet, their performance and reliability varies between different studies. The organizers of the first fastMRI challenge found that SSIM performed consistently with radiological evaluation \cite{Knoll_2020}. For the second fastMRI challenge, however, SSIM failed to detect hallucinations by numerous top-performing models \cite{Muckley_2021}. Additionally, two recent studies on the correlation of IQMs with radiological evaluation have reported SSIM and PSNR to perform worse than other reference-based metrics, e.g. feature similarity index (FSIM) and visual information fidelity (VIF) \cite{Mason_2020,Kastryulin_2023}. SSIM has also been shown to be less sensitive to simulated motion than e.g. VIF \cite{Terpstra_2024}. However, motion artifacts are complex, simulations often too simplistic \cite{Spieker_2024}, and none of these studies used real-motion data in their evaluation.

Alternatively, perceptual metrics based on deep features have been increasingly used in the computer vision and medical imaging community as an alternative to traditional IQMs \cite{Zhang_2018,Adamson_2023,miao2008quantitative}. Yet, they have not been comprehensively evaluated for medical imaging in general \cite{breger2024study}, nor for MR motion correction in particular. Moreover, all reference-based IQMs rely on a high-quality reference image. On the one hand, “hidden noise” in such reference images might influence metric values and lead to suboptimal ranking of different reconstructions \cite{Wang_2024}. On the other hand, in some scenarios - like prospective clinical studies or dynamic imaging - a ground-truth image might not be available at all. For these cases, quality evaluation relies on reference-free metrics. However, their development is challenging \cite{Chow_2016}, and they have shown less consistent correlation with radiological scores than reference-based metrics \cite{Eichhorn_2022,Marchetto_2024}.

In this work, we aim to assess the performance of commonly used reference-based and reference-free metrics in evaluating motion correction methods for research settings. We extend our previous evaluation of IQMs \cite{Eichhorn_2022, Marchetto_2024} with recent advances (VIF and perceptual image quality metric). Rather than being complete and comprehensive, our selection of IQMs focuses on the most relevant and commonly used metrics in the field of MR motion correction, as those offer a higher interpretability and acceptance in the community. We perform our evaluation on two unique datasets with real motion artifacts \cite{ganzdatasetOpenNeuro,marchetto2023robust}, which to the best of our knowledge has not been used for the analysis of IQMs so far. Further, we analyze the effect of common pre-processing steps on the IQMs, and their correlation with radiological assessment. The findings of our study might serve as recommendations for a reliable usage of IQMs in future studies.

\section{Methods}

\subsection{Image quality metrics}
In this study, we adopted ten IQMs: five reference-based and five reference-free metrics. The selection was made based on the metrics' popularity within the MR community\cite{Spieker_2024,Mason_2020,mcgee2000image}, code availability (when possible), and findings presented in the existing literature. We here report a list of the adopted metrics and provide the metrics' definitions in Table~\ref{tab:metrics}. For further implementation details we refer the reader to each reference as well as to our \href{https://github.com/melanieganz/ImageQualityMetricsMRI}{GitHub} repository.

\begin{table}[h]
    \caption{Definitions of image quality metrics.\label{tab:metrics}}
    \begin{tabular*}{\textwidth}{@{\extracolsep\fill}p{1.2cm}p{0.5cm}p{5.1cm}p{1.5cm}p{2cm}@{\extracolsep\fill}}
    \toprule
     & \textbf{Metric} & \textbf{Definition}  & \textbf{Values \newline ($\uparrow$ image quality)} & \textbf{Required image \newline value range}  \\
    \midrule 
     \textbf{Reference-based} & SSIM &  $\frac{1}{\lvert \mathcal{M} \rvert}\sum_{m,\hat{m} \in \mathcal{M}}\frac{(2\mu_m\mu_{\hat{m}}+c_1)(2\sigma_{m\hat{m}}+c_2)}{(\mu_m^2 + \mu_{\hat{m}}^2+c_1)(\sigma_m^2 + \sigma_{\hat{m}}^2+c_2)}$ & $\uparrow, \ limit: 1$  & - \\[0.3cm] 
    & PSNR & $10 \log_{10}\frac{\text{max}(\hat{x})^2}{\frac{1}{IJ}\sum_{i=1,j=1}^{I,J} (x_{ij}-\hat{x}_{ij})^2} $  &  $\uparrow$  & - \\[0.3cm]
    & FSIM &  \textit{Due to the complexity, please refer to Appendix~\ref{appendix:A}.} &   $\uparrow, \ limit: 1$   &  [0, 255] or [0, 1] \\[0.3cm]
    & VIF &  \textit{Due to the complexity, please refer to Appendix~\ref{appendix:A}.}  &  $\uparrow$ & [0, 255] or [0, 1] \\[0.3cm]
    & LPIPS &  $\textbf{d}(\mathcal{F}(x), \mathcal{F}(\hat{x}))$ &  $\downarrow, \ limit: 0$ & [-1, 1] \\[0.3cm]
    \midrule
    \textbf{Reference-free} & TG & $\frac{1}{IJ}\sum_{i=1,j=1}^{I,J} g_{i,j}^2 $   &  $\uparrow$ & - \\[0.3cm] 
    & AES & $\frac{\sqrt{\sum_{i,j} E(x_{i,j})g_{i,j}^2}}{\sum_{i,j} E(x_{i,j})}$  &  $\uparrow$ & -  \\[0.3cm]
    & NGS & $(\frac{g_{i,j}}{\sum_{i,j} g_{i,j}})^2$  &  $\uparrow$ & - \\[0.3cm]
    & IE &  $ -\sum_{i,j} y_{i,j} \ln(y_{i,j})$ with $y_{i,j} = \frac{x_{i,j}}{\sqrt{\sum x_{i,j}^2}}$ & $\downarrow$ & - \\[0.3cm]
    & GE &  $ -\sum_{i,j} z_{i,j} \ln(z_{i,j})$ with $z_{i,j} = \frac{g_{i,j}}{\sqrt{\sum g_{i,j}^2}}$ & $\downarrow$  & - \\[0.3cm]
    \botrule
    \end{tabular*}

    \begin{tablenotes}
    \item $x$: image to be evaluated; $\hat{x}$: reference image; 
    $m$/$\hat{m}$: patch of $x$/$\hat{x}$, $\mu$: mean value, $\sigma$: standard deviation, $c_1/c_2\propto L^2$: variables proportional to dynamic range $L$;
    $\mathbf{d}$: distance measure; $\mathcal{F}$: features extracted with pre-trained neural network;
    $g_{i,j} = \sqrt{(\nabla_x x_{ij})^2 + (\nabla_y x_{ij})^2}$: gradient magnitude;
    $E(x)$: binary mask of edges of $x$;
    $\uparrow$: metric value increases as image quality increases; 
    $\downarrow$: metric value decreases as image quality increases.
    \end{tablenotes}
\end{table}

\subsubsection*{\textbf{Reference-based metrics}}
\begin{itemize}
\item \textbf{Structural Similarity Index Measure (SSIM)} \cite{wang2004image} measures the similarity between two images by evaluating luminance, contrast, and structure similarity. It provides a value between -1 and 1, where 1 indicates perfect similarity.
\item \textbf{Peak Signal-to-Noise Ratio (PSNR)} \cite{hore2010image} measures the ratio between the maximum possible power of a signal and the power of corrupting noise. It is expressed in decibels (dB), with higher values indicating a better image quality.
\item \textbf{Feature Similarity Index Measure (FSIM)} \cite{zhang2011fsim} calculates the image similarity using the phase congruency on the frequency representation of the magnitude image, which detects edge similarities. High phase congruency values in Fourier components identify sharp light-dark transitions, perceived as edges. Gradient magnitude, added to account for contrast invariance, enhances the metric. FSIM ranges from 0 to 1, with 1 indicating identical images.
\item \textbf{Visual Information Fidelity (VIF)} \cite{sheikh2006image} is a metric based on natural scene statistics, designed to evaluate the quality of images based on the information they convey to the human visual system. One appealing feature of VIF is its ability to measure improvements in image quality compared to the reference image, which is indicated by a value greater than 1.  
\item \textbf{Perceptual Image Patch Similarity (LPIPS)} \cite{Zhang_2018} measures the distance between features extracted from two images with a pre-trained convolutional neural network. LPIPS is 0 for identical images and increases with decreasing similarity. 
\end{itemize}

\subsubsection*{\textbf{Reference-free metrics}}
\begin{itemize}
\item \textbf{Tenengrad (TG)} \cite{kecskemeti2018robust} is a gradient-based metric commonly used to assess image sharpness or focus. It measures the intensity of edges by averaging gradient magnitudes across the image. Higher values indicate sharper images with prominent edges.  
\item \textbf{Average Edge Strength (AES)} \cite{pannetier2016quantitative,zaca2018method} is a similar gradient-based metric. It is designed to quantify the overall edge content in an image by calculating the average gradient magnitude across detected edges. Higher values indicate more pronounced edges, typically associated with sharper images.
\item \textbf{Normalized Gradient Square (NGS)} \cite{mcgee2000image} is another gradient-based metric, used to assess image sharpness. It is a normalized version of TG, providing a relative measure of image focus.
\item \textbf{Image Entropy (IE)} \cite{Atkinson_1997,mcgee2000image} is a statistical metric that quantifies the amount of randomness in an image by analyzing the distribution of pixel intensities. Lower entropy values indicate more uniform, ordered pixel intensities, which are typically associated with higher image quality, such as sharper or less noisy images. We follow the implementation of Atkinson et al.\cite{Atkinson_1997}.
\item \textbf{Gradient Entropy (GE)} \cite{mcgee2000image} combines gradient- and entropy-based evaluation. It calculates the entropy of the gradient magnitudes of an image and provides a measure of the randomness or complexity of the image's edge structures. Lower values typically indicate more structured and concentrated edges, reflecting higher image quality.
\end{itemize}

\subsection{Data acquisition}
In this study, we utilized two different datasets: 
First, a publicly available dataset acquired at the Neurobiology Research Unit (\textbf{NRU}, Copenhagen, Denmark)\footnote{\url{https://openneuro.org/datasets/ds004332/versions/1.1.3}}\cite{ganzdatasetOpenNeuro}. This dataset includes 3D $T_1$ MP-RAGE, 3D $T_2$ FLAIR, 2D $T_1$ STIR, and 2D $T_2$ TSE acquisitions with instructed head nodding and  shaking motion from 22 healthy participants. 
Each sequence was acquired with and without voluntary motion, as well as with and without prospective motion correction. The acquisition without motion and without motion correction served as reference image.  
Second, a private dataset acquired at the Cardiff University Brain Research Imaging Centre (\textbf{CUBRIC}, Cardiff, UK)\cite{marchetto2023robust}. This dataset consisted of solely MP-RAGE images from 9 healthy participants. Reference images were available for each subject, and the dataset comprised of acquisitions with and without voluntary motion. The motion types included nodding, continuous circular head movements, and "step-wise" motion. Retrospective motion correction was applied to the whole dataset, while uncorrected images remain available.

Both datasets were acquired on 3\,T Prisma MRI scanners (Siemens Healthineers, Erlangen, Germany). Further information regarding acquisition details, types of voluntary motion and motion correction methods for both datasets can be found in \cite{ganzdatasetOpenNeuro,marchetto2023robust}.

\subsection{Pre-processing} \label{sec:preproc}
\begin{figure}[t]
\centerline{\includegraphics[width=0.9\linewidth]{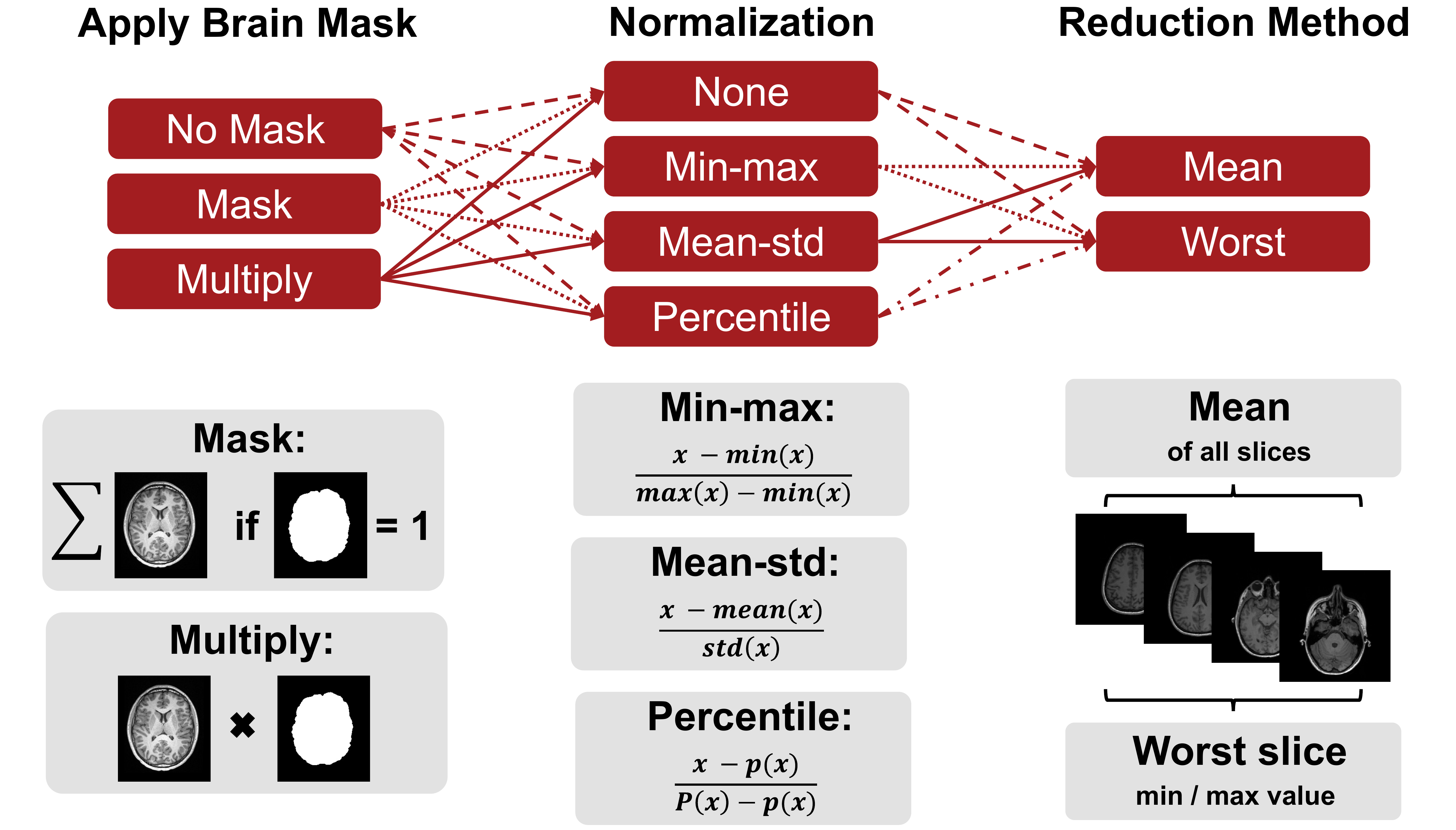}}
\caption{Different pre-processing choices are involved for calculating IQMs. We vary three of the common pre-processing steps, namely masking, normalization and reduction method of the IQM values.  The brain mask was either neglected, multiplied to the images or the metric was only evaluated within  brain mask voxels. Images were either not normalized or normalized with min-max, mean-std or percentile normalization (except for FSIM, VIF, and LPIPS which require specific image values as shown in \ref{tab:metrics}). IQM values across slices were reduced by calculating the mean value or taking the worst value of all slices (min/max depending on IQM).}
\label{fig:preproc_choices}
\end{figure}
Our pre-processing pipeline comprised five different steps to estimate the IQMs: skull-stripping, alignment, masking, normalization and the method used to reduce a set of IQM values across slices to a single value. Skull-stripping was performed on the reference MP-RAGE images using the Brain Extraction Tool (BET) \cite{smith2002fast} (further parameters -R -f 0.4 -m). For each sequence, the non-reference images were co-registered with the respective reference image. The brain mask extracted from the MP-RAGE reference was co-registered to the reference image of the remaining sequences (3D FLAIR, 2D TSE, 2D TIRM), to ensure brain masks to be in the same space as the corresponding sequence. Both alignments were performed using the rigid registration option in FLIRT (FMRIB's Linear Image Registration Tool) \cite{jenkinson2002improved}. To avoid inconsistencies with peripheral slices, only slices containing at least 10\% brain voxels were included in the pre-processing and subsequent analysis. 

While we fixed these first two steps with respect to the tooling used, we varied the masking, normalization and reduction method of the slice-wise IQM values into a single value (mean or worst slice), as illustrated in Fig. \ref{fig:preproc_choices}. First, the images were either (i) not masked, (ii) masked directly (where only intensities inside the brain were used during metric calculation) or (iii) masked through multiplication with the brain mask (which effectively zeroes out the background of the image). Second, the intensities were normalized volume-wise following (i) a min-max, (ii) a mean divided by standard deviation, or (iii) a percentile ($1^{st}$ / $99.9^{th}$) normalization approach. Alternatively, (iv) the intensities were not normalized at all\footnote{Note that FSIM, VIF and LPIPS require a specific image value range (see Table~\ref{tab:metrics}). These metrics can only be calculated for min-max and percentile normalization and require an additional rescaling to the respective value ranges after normalization.}.   
Third, the IQM values were computed for each slice and the final metric value was determined as either the mean or the worst value among all slices (min/max depending on IQM).


\subsection{Image quality assessment}\label{sec:rad_scores}
\begin{figure}[t]
\centerline{\includegraphics[width=0.8\linewidth]{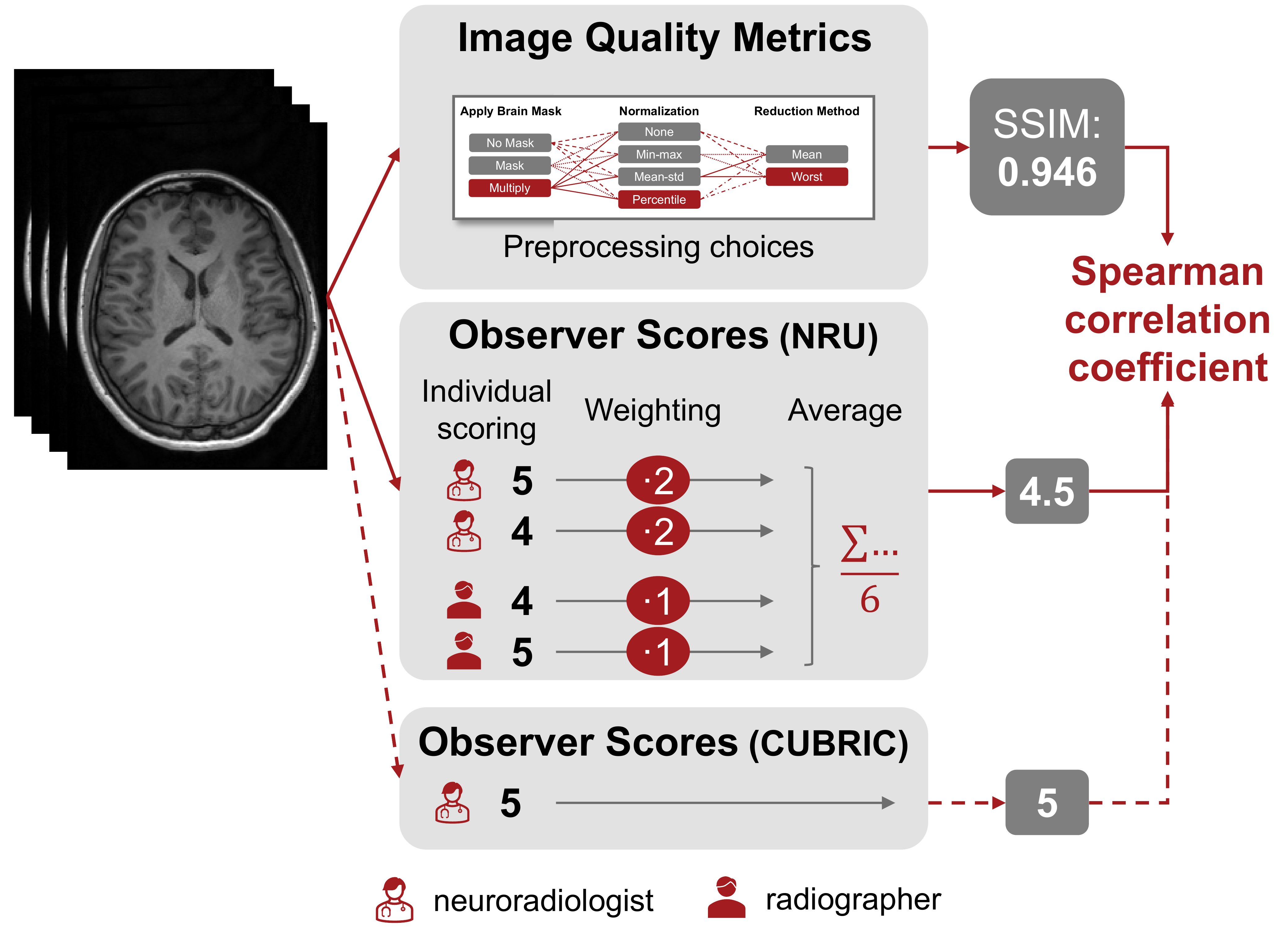}}
\caption{Overview of the correlation analysis between image quality metrics and observer scores. Each 3D image volume was evaluated by one neuroradiologist for the CUBRIC dataset and by two neuroradiologists and two radiographers for the NRU dataset. For the latter, the scores were averaged with double weight on the more experienced neuroradiologists. IQMs were computed with various preprocessing choices (compare Fig.~\ref{fig:preproc_choices}), as illustrated exemplary for SSIM. IQM values and observer scores of all images were then used to calculate the Spearman correlation coefficient to measure the agreement between IQMs and observers.}
\label{fig:overview_comparisons}
\end{figure}

As illustrated in Fig.~\ref{fig:overview_comparisons}, the anonymized images from the NRU dataset were evaluated by two neuroradiologists with over 10 years of experience in reading MR images (N.S. and S.S.) and two recently graduated radiographers (M.R.R. and B.P.). For the CUBRIC dataset, ratings were performed by one of the experienced neuroradiologists (S.S.).
Because of the different level of experience in evaluating medical images, we averaged the scores with a double weight on the radiologists. The image assessment was performed using a 1-5 Likert scale \cite{sullivan2013analyzing}, with 5 representing a perfect image (without artifacts) and 1 a completely non-diagnostic image. Both radiologists and radiographers were instructed to score the images based on the worst slice within the volume. The intra-variability between evaluators was assessed using the Krippendorff's alpha coefficient, which ranges from 0 (no agreement) to 1 (perfect agreement), with values above 0.8 typically considered indicative of good reliability.\\
The correlation between the IQM values and the scores given by the evaluators was estimated using the Spearman rank correlation coefficient \cite{spearman1987proof}. While the Pearson correlation coefficient uses a linear function, the Spearman correlation coefficient applies a monotonic function to measure strength and direction of the relationship between the two variables, which are also not required to be normally distributed \cite{schober2018correlation}. The Spearman rank correlation coefficient spans between -1 and 1, representing a perfectly monotonic negative and positive relationship between the two variables, respectively. Spearman correlation magnitudes above 0.7 indicate strong correlations \cite{schober2018correlation}.

\section{Results}

\subsection{Validity of observer scores}
First, we tested the validity of the observer scores for the NRU dataset. The Krippendorff's alpha coefficient shows good agreement between the observers in case of the MP-RAGE sequence, with a value of 0.82. For the $T_{2}$ TSE, $T_{2}$ FLAIR and $T_{1}$ STIR images the evaluators displayed moderate agreement with values of 0.78, 0.70 and 0.71 respectively.

\begin{figure}
\centering
\includegraphics[width=0.8\textwidth]{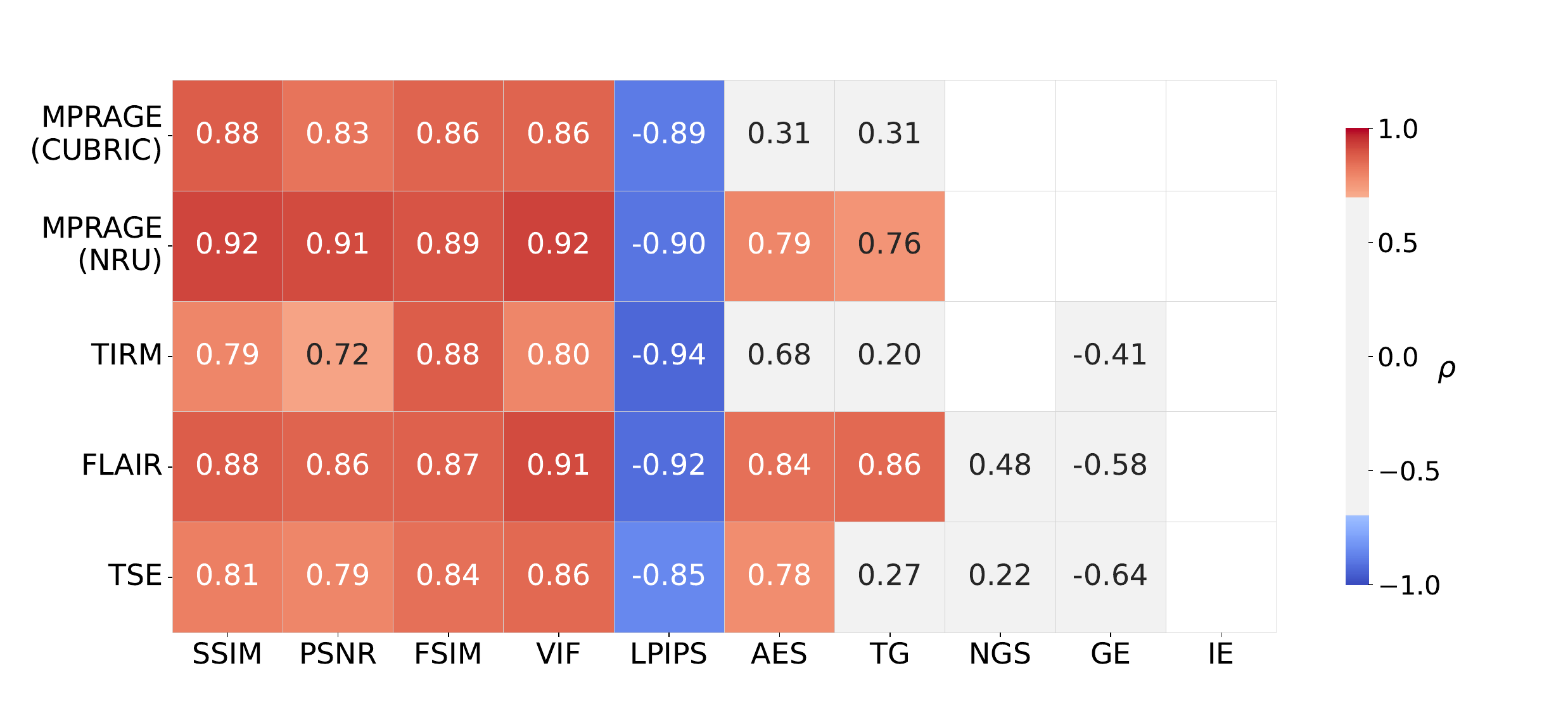}
\put(-330,110){\textbf{(A)}}
\vspace{-1.0em} 
\includegraphics[width=0.8\textwidth]{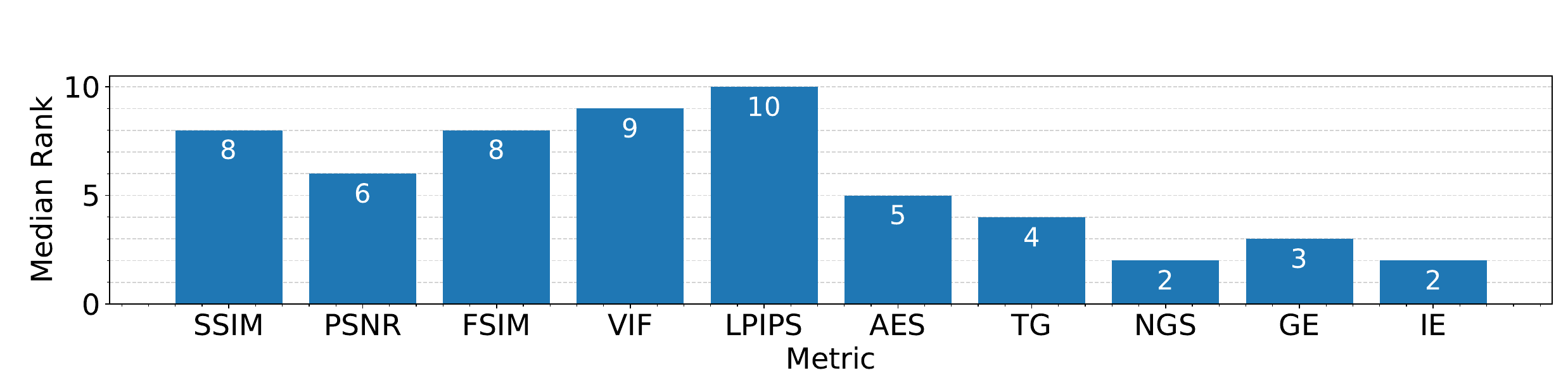}
\put(-330,50){\textbf{(B)}}
\caption{(A) Spearman correlation coefficient $\rho$ between IQMs (x-axis) and observer scores for the four sequences of the NRU dataset (y-axis). Values are provided for statistically significant correlations (p-value $< 0.05$) and values corresponding to a strong correlation ($ \mid \rho \mid > 0.6$) are colored in blue and red. The metrics were calculated with the pre-processing settings \{\textit{Multiply, Percentile, Worst}\}. (B)~Median rank of each IQM, resulting from ranking the absolute values of the correlation coefficients for each sequence and taking the median across sequences.}
\label{fig:iqms_main}
\end{figure}

\begin{figure}
\centering
\includegraphics[width=1\linewidth]{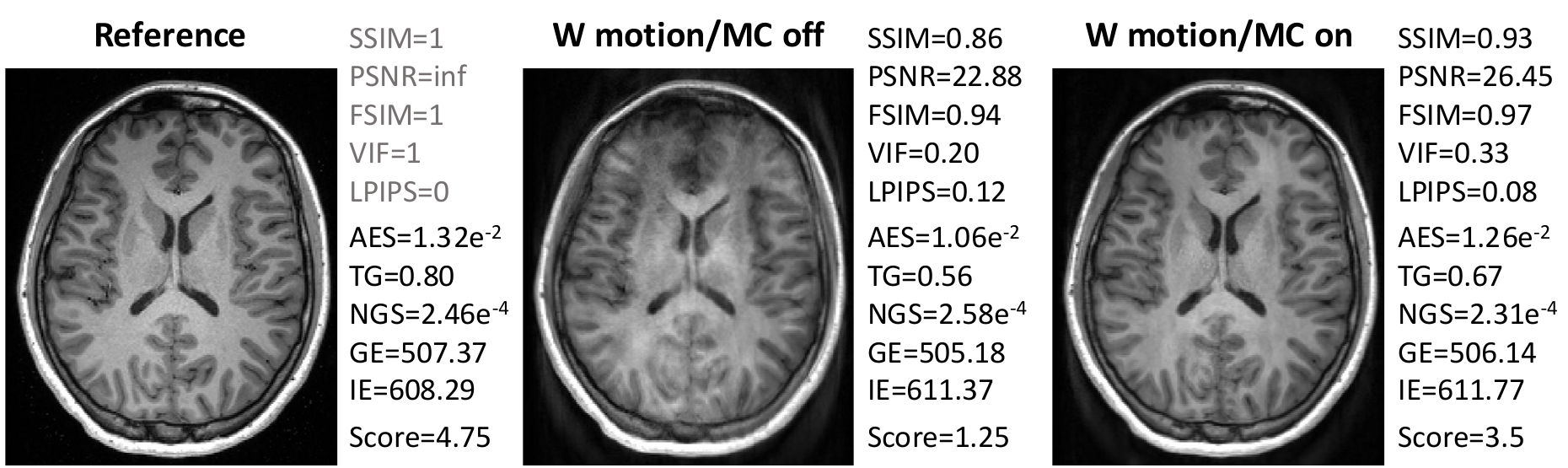}
\vspace{-1em} 
\caption{Three examples of MP-RAGE images from one subject. The reference image was acquired without voluntary motion and without motion correction, while the other two examples were acquired with voluntary motion (nodding/shaking) and with/without motion correction. Image quality metrics are reported, alongside with the average observers' evaluation scores ("Score"). Examples for $T_2$ FLAIR, $T_1$ TIRM and $T_2$ TSE are shown in Fig.\,S2. 
The reference-based IQMS (SSIM, PSNR, FSIM, VIF, LPIPS) and reference-free IQMS (AES, TG, NGS, GE, IE) are shown to the right of the images. For the reference image, the reference-based IQMS are calculated on itself and colored in light-gray. 
}
\label{fig:comp_mprage}
\end{figure}

\subsection{Correlation of IQMs with observer scores across MR sequences}
The correlation of the analyzed IQMs with observer scores for both datasets is compared in Fig.~\ref{fig:iqms_main} for the pre-processing settings \{\textit{Multiply, Percentile, Worst}\}. All reference-based IQMs show a strong correlation with radiological assessment, with small variations in their relative performance for different MR sequences. Among the reference-free IQMs, AES and TG perform best, but correlations are not as strong and less consistent across sequences and datasets as for reference-based IQMs. Fig.~\ref{fig:comp_mprage} compares three MP-RAGE example images from one subject with varying levels of motion, showing image quality metrics alongside the average evaluation scores from the observers.

\begin{figure}
\centerline{\includegraphics[width=0.95\linewidth]{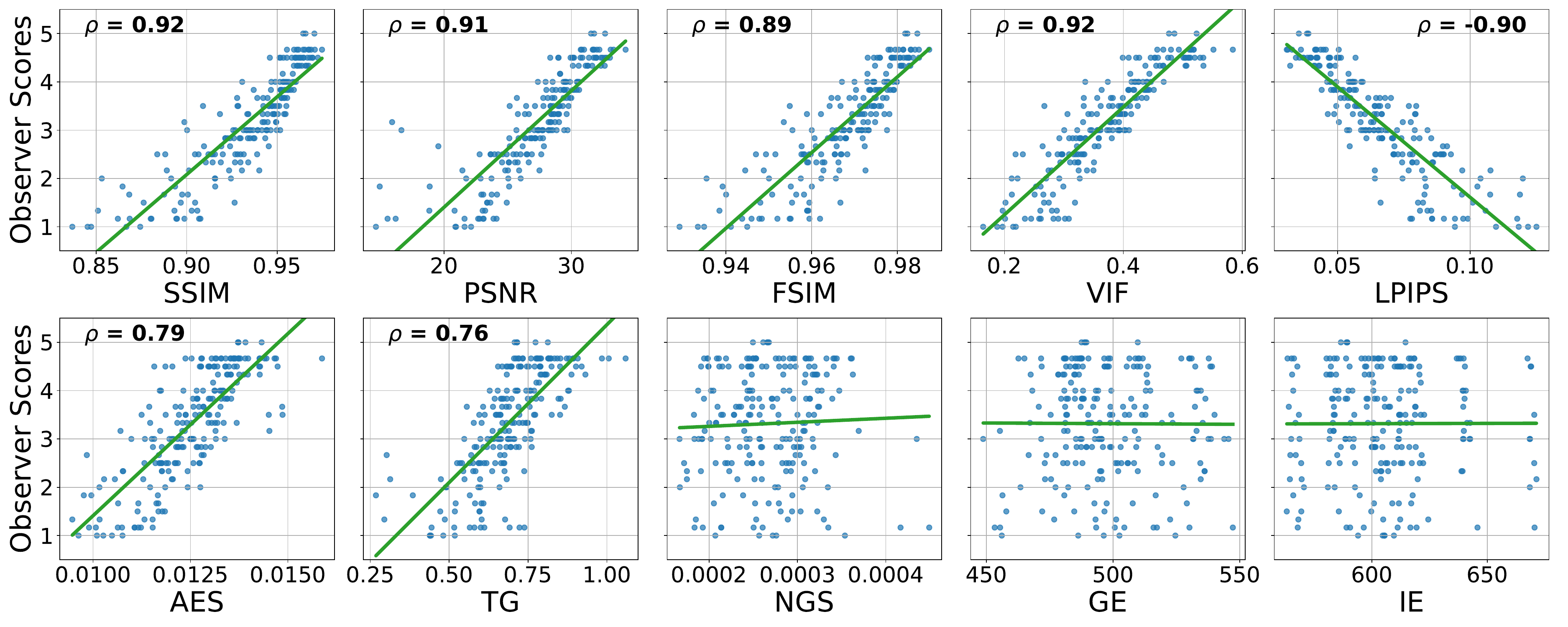}}
\caption{Scatter plots visualizing the distribution of metrics values against observer scores. Each blue dot represents one MP-RAGE image volume from the NRU dataset and the corresponding regression line is shown green. For statistically significant correlations (p-value $< 0.05$), the corresponding Spearman correlation coefficient is provided on top of the plot. The metrics were calculated with the pre-processing settings \{\textit{Multiply, Percentile, Worst}\}. Non-integer observer scores result from averaging the scores across the four raters.}
\label{fig:scatter_plots_mprage}
\end{figure}

To provide further context on these abstract correlation values, Fig.~\ref{fig:scatter_plots_mprage} shows the scatter plots of metric values against observer scores for the MP-RAGE sequence of the NRU dataset. Plots for the other sequences can be found in the Supplementary Information (Fig. S1).

\subsection{Influence of implementation decisions}
To test the robustness of the IQMs towards standard implementation variations, we compared the strength of correlation for different pre-processing settings. We display the results for the MP-RAGE sequence (of both NRU and CUBRIC datasets) in Fig.~\ref{fig:preproc_mprage}, while the plots for the other sequences can be found in the Supplementary Information (Figs. S2, S3 and S4).
We did not observe a significant difference in the correlation coefficients for different slice reduction methods, i.e. whether the metric value of the worst slice is chosen or the mean of all slices is calculated. However, with respect to different normalization methods, we observed inconsistent correlation results, particularly for PSNR, AES and TG. Percentile normalization performed best over all metrics. Further, with respect to the brain mask application, we did not notice substantial differences between masking metric values or multiplying the images with the mask, but correlations dropped significantly when no brain mask was applied at all.

\begin{figure}
    \centering
    \includegraphics[width=0.46\textwidth]{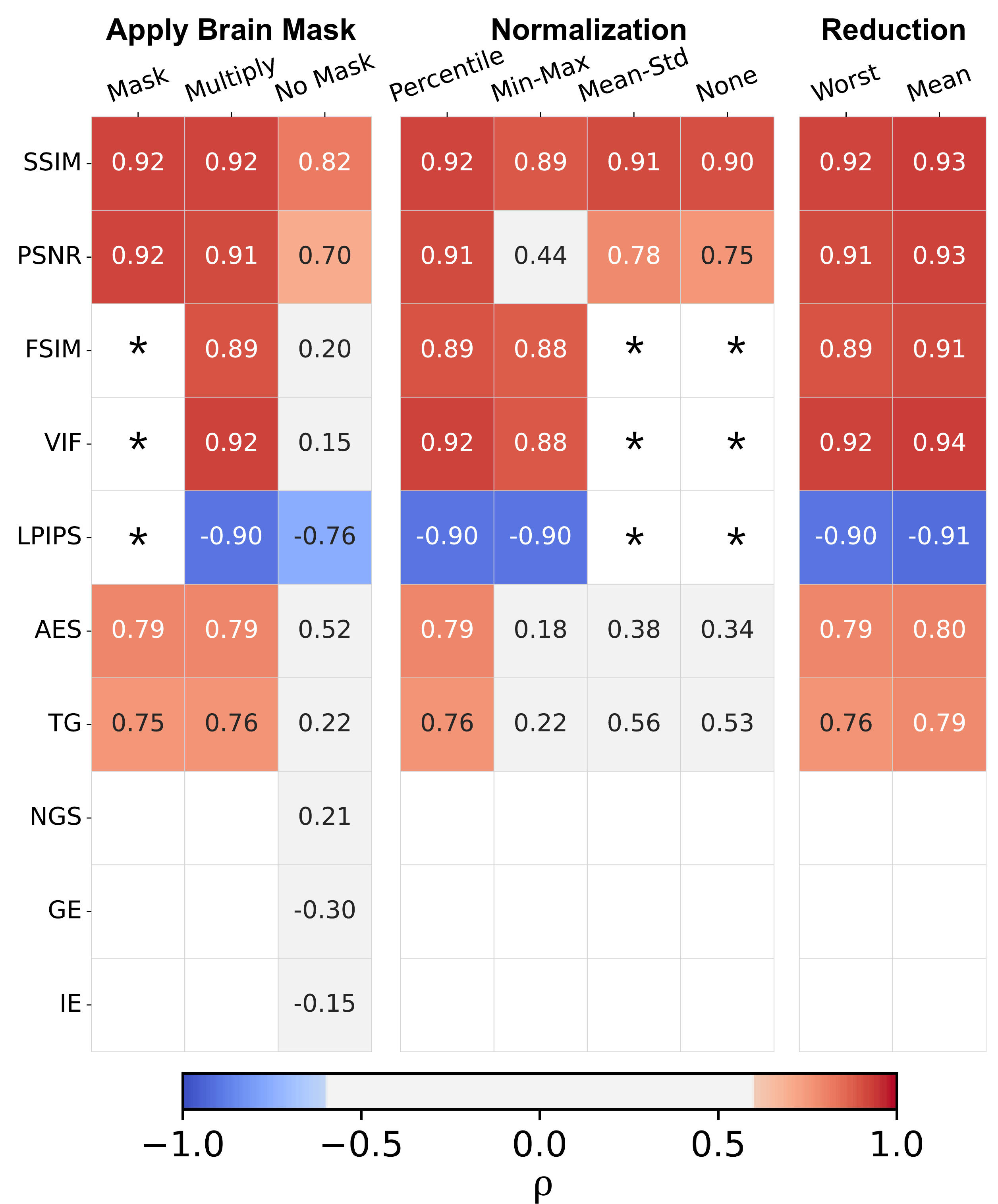}
    \hspace{0.01\textwidth}
    \includegraphics[width=0.46\textwidth]{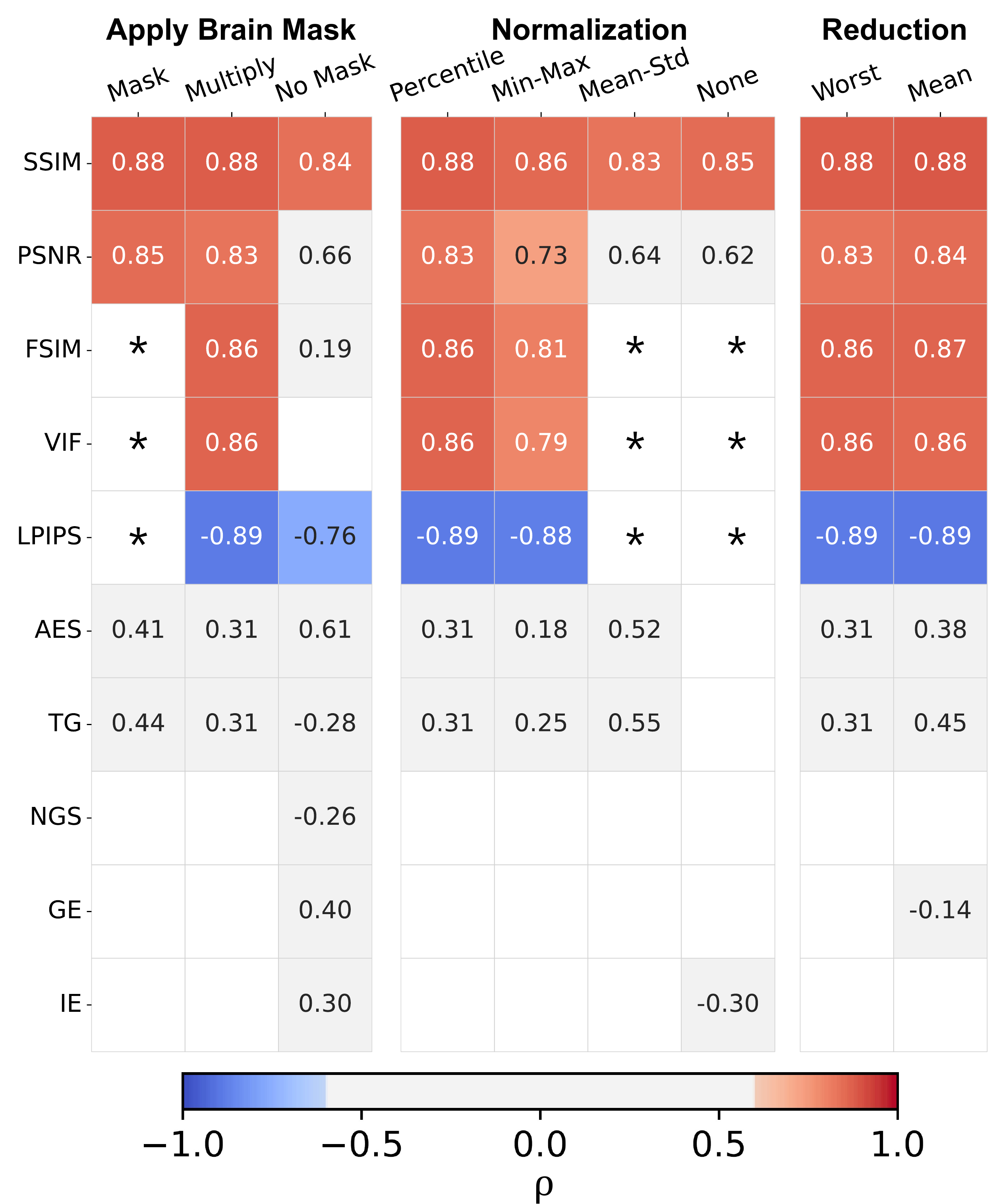}
    \put(-358,195){\textbf{(A)}}
    \put(-175,195){\textbf{(B)}}
    \caption{Overview on the effect of pre-processing implementations in the correlation between IQM and observers' scores on the MP-RAGEs from the NRU (A) and the CUBRIC dataset (B). We compare the different options for each pro-processing choice individually, while keeping the other two pro-processing settings at the standard \{\textit{Multiply, Percentile, Worst}\}.
    The table only shows statistically significant correlations ($p<0.05$), leaving the box empty if this requirement is not fulfilled. We indicated with a "$*$" values for FSIM, VIF and LPIPS which are not available in case of normalization using "Mean-Std" and "None", as they require a specific range of values (see Table \ref{tab:metrics}). Similarly, these values are unavailable with the "Mask" setting, as the metrics are computed across the entire matrix. Overall, we found that the correlations with reference-based metrics are more consistent compared to the reference-free metrics, which largely display weak correlation with the observer's evaluations. The pre-processing steps that mostly affect the correlation values are: not applying a brain mask ("No Mask"), applying no normalization ("None") or rescaling using the "Mean-Std" method.}
    \label{fig:preproc_mprage}
\end{figure}

\begin{figure}
\centering
\includegraphics[width=1\linewidth]{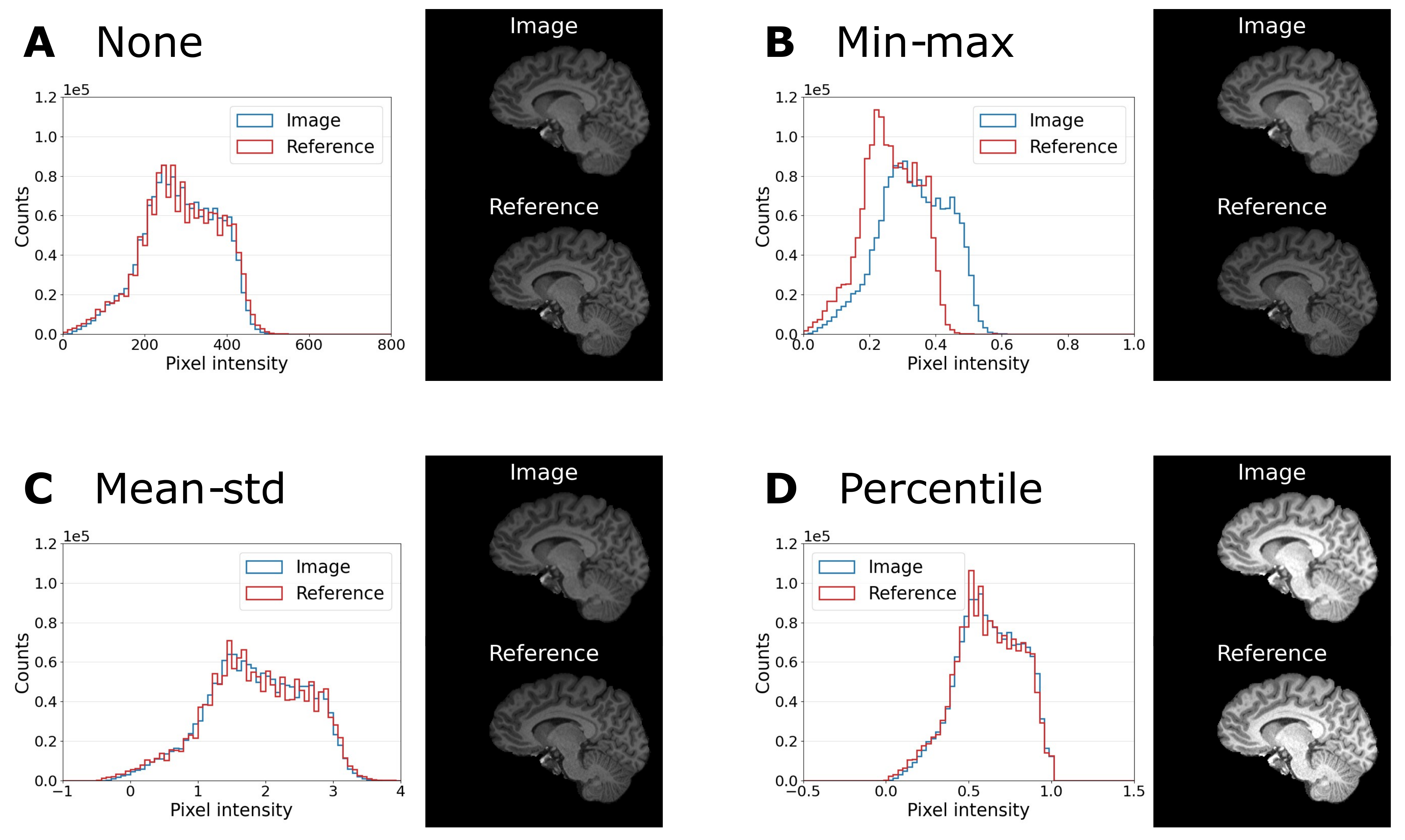}
\caption{Intensity distributions of one example MP-RAGE image (blue) and its reference (green) for the normalization settings (A) “None”, (B) “Min-max”, (C) “Mean-std” and (D) “Percentile”. The analysis is performed only within the brain mask. Example slices of image and reference (same intensity window) are shown next to the histograms. Min-max normalization is impacted by large outlier values and leads to a mismatch of intensity values in image and reference.}
\label{fig:hist_norm}
\end{figure}

\section{Discussion}
We have assessed the correlation of image quality metrics with radiological evaluation under various pre-processing settings for two datasets with real motion artifacts. Our results confirm that reference-based IQMs exhibit consistently strong correlations with radiological assessments. Among the reference-free IQMs, only AES and TG correlate consistently with observer scores across all four MR sequences. 
Pre-processing choices have a varying influence on the stability of the correlations. We have found that the robustness of the IQM estimation was largely unaffected by variations in slice reduction methods. Normalization techniques, in contrast, significantly influenced correlation strength, with percentile normalization outperforming others. Furthermore, the use of brain masks proved essential, as the absence of a mask led to a substantial drop in correlation. 

We have investigated the causes of the large variations due to pre-processing choices. To explain the influence of normalization, we have compared the distribution of pixel intensities for the four different normalization methods for one example MP-RAGE image and its corresponding reference in Fig.~\ref{fig:hist_norm}. This illustrates that min-max normalization is impacted by large outlier values, while mean-std and percentile normalization better match the histograms of the image and its reference. But additional methods of normalization could be considered as well, such as slice-wise normalization. The influence of applying the brain mask is the most extensive, but also easy to understand. In our brain application the background covers more than 60\% of the image. Therefore, using the background when estimating IQMs will naturally bias the results since they will be largely driven by the background. This is also not desirable when one wants to assess image quality, since in a clinical evaluation the focus obviously lies on the image and not the background. This issue might be specific to brain imaging. Other application areas, e.g. cardiac or abdominal imaging, have a much larger portion of the image that contains no background and therefore masking is of lesser importance, and should be part of future investigation. 

The observer scoring was performed on the worst slice of the image volume, based on discussions with the neuroradiologists, since when they are looking for brain abnormalities, e.g. epilepsy lesions or bleeds, they examine all slices of an image. Even if only some of the slices are degraded a proper diagnosis might not be possible. We tried to address this by comparing the mean vs. worst reduction method for the IQM values across slices and we have found that the IQM estimation was largely unaffected by variations in slice reduction methods.

The comparison between the MP-RAGE datasets acquired at different institutions shows consistent smaller correlation values for the CUBRIC dataset compared to NRU (Fig.~\ref{fig:preproc_mprage}). This discrepancy might be attributed to the larger variety of motion patterns performed during the acquisition of the CUBRIC dataset compared to NRU, as well as to the limited quality evaluation (only one experienced radiologist for the CUBRIC dataset versus two radiologists and two technicians for the NRU dataset).

Finally, in this work we focused on the process of calculating IQMs. But additional pre-processing steps, such as different brain mask extraction methods (FreeSurfer vs. BET vs. SPM) might also influence the results. If the masking completely fails and does not cover the whole region of interest, then no reliable IQM can be calculated. If it fails to remove all of the background due to e.g. excessive ghosting, then the IQM would be affected, but probably lead to a similar result than not removing the background. Hence, this should be assessed in future work. Moreover, mis-registration can substantially affect the accuracy of image quality metric evaluations. Therefore, we suggest to inspect the quality of the registration, as different registration settings (FreeSurfer vs. FSL vs. SPM) have not been tested yet. In the intermediate, we strongly recommend to clearly describe all pre-processing steps, including which brain mask extraction and registration tool was utilized, and to share the analysis code openly.

Given the above, we cannot recommend a single IQM, but instead advocate for using a set of metrics to reflect different properties. But in general, we see that if reference-based metrics are possible to compute, those perform better and are closer to radiological assessments and therefore preferable.

\subsection{Limitations}
Our current analysis is based on data acquired for research purposes that included a separate 'still' reference scan. The reason for this is that we wanted to assess reference-based and reference-free IQMs and that IQMs are currently largely used to evaluate image quality, e.g. for sequence development in the MR physics community. But of course, IQMs would also be desirable to be used in a clinical setting in order to provide inline quality assessment of MRI scans to reduce re-scans. Hence, it is desirable to assess if there exist any reference-free IQMs that correlate well with radiological assessments and to check the influence of pre-processing choices on clinical data as well. Some of the pre-processing choices might need to be adapted when they are applied to clinical data especially to 2D sequences with varying coverage. Moreover, our datasets are limited to 3T acquisitions: future studies should therefore evaluate the performance of image quality metrics at different field strengths and sub-millimeter resolutions \cite{bazin2020sharpness}.
Finally, the presented IQMs are not proper metrics in the mathematical sense and will therefore always vary in values. Hence, a direct comparison of metric values between studies, is only possible if exactly the same metric implementation is used on exactly the same dataset. This precludes us from directly comparing studies of e.g. different motion correction methods and points in the direction of necessitating data sharing of standard datasets for methods testing.

\subsection{Outlook}
To bridge the gap between reference-free and reference-based IQMs, future developments could focus on distribution-based metrics and learning-based approaches. In particular, approaches are favorable that do not require matched reference images but learn statistical properties of motion-free and motion-corrupted images and thus mimic how radiologists assess image quality. Initial work on automated image quality assessment without reference images has demonstrated the potential of such approaches for specific sequences \cite{mortamet2009automatic,pizarro2016automated,kustner2018automated}. With the growth of large-scale datasets and computational resources, more powerful models can be trained in the future \cite{Ecker_2024}, potentially enabling automated image quality assessment to become more robust and generalizable across applications.

\section{Conclusion}
In our study, we have evaluated the correlation between image quality metrics (IQM) and radiological scores, and have shown how different pre-processing steps can strongly affect the correlation between IQMs and radiological assessment. Overall, we have found that reference-based IQMs show consistently stronger correlations than reference-free metrics across different datasets and image contrasts. Most importantly, our findings underscore the importance of pre-processing choices in IQM-based quality assessment, as well as the need for sharing detailed documentation, in the spirit of reproducible research.

\backmatter

\bmhead{Supplementary information}

Electronic supplementary information is available.

\bmhead{Acknowledgements}

We thank the two experienced neuroradiologists, Assistant professor of Radiology Nitesh Shekhrajka from the University of Iowa hospitals and Clinics in the United States and Consultant Neuroradiologist Stefan Schwarz from the Cardiff University Hospitals in the United Kingdom, for performing the observer quality scoring. We also thank radiographers Martin Riis Rassing and Bianca Pedersen who contributed with observer quality scoring.

Additionally we would like to acknowledge the following funding sources:

\section*{Declarations}
\subsection*{Funding}
 Elisa Marchetto's work was performed under the rubric of the Center for Advanced Imaging Innovation and Research (\href{www.cai2r.net}{$CAI^2R$}), an NIBIB National Center for Biomedical Imaging and Bioengineering (NIH P41 EB017183). Hannah Eichhorn is partially supported by the Helmholtz Association under the joint research school ”Munich School for Data Science - MUDS”. Melanie Ganz-Benjaminsen was supported by the Elsass Foundation (18-3-0147).
\subsection*{Conflict of interest}
The authors declare no potential conflict of interests.

\subsection*{Ethics approval and consent to participate}
Ethical approval for this study was obtained from Cardiff University School of Psychology Ethics Committee board. The nine healthy participants were recruited to take part in the study between the 19th of October 2020 and the 31st of March 2021.  The Copenhagen study was approved by the local scientific ethics committee and the Danish Data Protection Agency prior to initiation (Cimbi database H-KF-2006-20). Written informed consent was obtained from all individual participants included in the study.

\subsection*{Data availability}
The dataset acquired at the Neurobiology Research Unit (Copenhagen, Denmark) is publicly available at \url{https://openneuro.org/datasets/ds004332/versions/1.1.3}\cite{ganzdatasetOpenNeuro}

\subsection*{Code availability }
The code for pre-processing and image quality metrics calculation was developed in Python 3.12.3 and is publicly available on \href{https://github.com/melanieganz/ImageQualityMetricsMRI}{GitHub}.

\subsection*{Authors’ Contribution}
Eichhorn and Marchetto contributed equally to this work.
Marchetto: Study conception and design; Acquisition of data; Analysis and interpretation of data; Drafting of manuscript; Critical revision.
Eichhorn: Study conception and design; Acquisition of data; Analysis and interpretation of data; Drafting of manuscript; Critical revision.
Gallichan: Analysis and interpretation of data; Critical revision.
Schnabel: Analysis and interpretation of data; Critical revision.
Ganz: Study conception and design; Acquisition of data; Analysis and interpretation of data; Drafting of manuscript; Critical revision.

\begin{appendices}

\section{Mathematical description}
\label{appendix:A}
This section is meant to provide some additional information regarding the FSIM \cite{zhang2011fsim} and VIF \cite{sheikh2006image} metrics reported in Table \ref{tab:metrics}. For additional information, please refer to the respective references. 

\subsection*{FSIM}
The Feature Similarity Index Measure (FSIM) is calculated by first computing the similarity measure between the two images wrt. the phase congruency (PC) \cite{Kovesi_1999} and the gradient magnitude (GM):
\begin{equation}
    S_{PC} = \frac{2PC_1(x)\cdot PC_2(x) + T_1}{2PC_1^2(x) + PC_2^2(x) + T_1}
    \label{eq:spc}
\end{equation}

\begin{equation}
    S_{GM} = \frac{2GM_1(x)\cdot GM_2(x) + T_2}{2GM_1^2(x) + GM_2^2(x) + T_2}
    \label{eq:gm}
\end{equation}
where $T_1$ and $T_2$ are two positive constant defined to increase the stability of the two metrics.\\
From Eq. \ref{eq:spc} and \ref{eq:gm} we can derive the similarity index $S_L$ as:
\begin{equation}
    S_{L} = [S_{PC}(x)]^\alpha \cdot [S_{GM}(x)]^\beta
    \label{eq:sl}
\end{equation}
with $\alpha$ and $\beta$ being weights to adjust the relative importance of the two terms. In this paper they were both kept at 1 as in the original implementation \cite{zhang2011fsim}.\\
Finally the FSIM index can be derived as:
\begin{equation}
    \mathit{FSIM} = \frac{\sum_{x\in\Omega}S_L(x) \cdot PC_m(x)}{\sum_{x\in\Omega}PC_m(x)}
\end{equation}
where $PC_m(x) = max(PC_1(x), PC_2(x))$.

\subsection*{VIF}
Visual Information Fidelity (VIF) quantifies the similarity between two images, here called test and reference images, capturing how well the reference's information is preserved. The approach consists on measuring the information fidelity across multiple scales (resolution) by applying a Gaussian filter. The VIF is then calculated as the ratio of the information conveyed by the test image to the information available in the reference image. The VIF is then computed as:
\begin{equation}
\mathit{VIF} = \log_{10} \left( 1 + \frac{g^2 \cdot \sigma_y^2}{\sigma_v^2 + \sigma_n^2} \right)
\end{equation}
\begin{equation}
  g = \frac{\sigma_{xy}}{\sigma_y^2 + \epsilon}  
\end{equation}
\begin{equation}
\sigma_v^2 = \sigma_x^2 - g \cdot \sigma_{xy}
\end{equation}
with $\sigma_x^2$, $\sigma_y^2$, and $\sigma_{xy}$ being the variances and covariance respectively of the test and reference images, and $\sigma_n^2$ the variance of the noise.
The overall VIF index is obtained by summing the contributions from all scales and normalizing them, resulting in a value within the interval \([0, 1]\), exceeding 1 for images with enhanced contrast.

\end{appendices}

\bibliography{Bibliography}

\end{document}